\documentclass[12pt]{article}
\usepackage[utf8]{inputenc}
\usepackage[english]{babel}
\usepackage{amsmath}
\usepackage{amsfonts}
\usepackage{amssymb}
\usepackage{makeidx}
\usepackage{graphicx}
\usepackage[ruled,vlined]{algorithm2e}
\usepackage{lmodern}
\usepackage{kpfonts}
\usepackage{natbib}
\usepackage{ctable}
\usepackage[left=2.5cm,right=2.5cm,top=2.5cm,bottom=2.5cm]{geometry}
\usepackage[]{geometry}
\usepackage{comment}
\usepackage{caption}
\usepackage{setspace}

\providecommand{\keywords}[1]
{
  \small	
  \textbf{\textit{Keywords---}} #1
}

\title{Optimized questionnaire item selection for tracking the progression of motor symptoms in Parkinson's disease}
\author{Karl Sigfrid$^{1}$, Ellinor Fackle-Fornius$^{1}$, Frank Miller$^{2}$\\[2mm]
$^1$Department of Statistics, Stockholm University, Sweden\\
$\mbox{}^2$Department of Computer and Information Science,\\Link\"oping University, Sweden}
\date{}

\begin{document}

\maketitle

%\doublespacing

\begin{abstract}
Long questionnaires increase the response burden for patients and healthcare workers. In the treatment of Parkinson's disease, the MDS-UPDRS questionnaire to track disease progression may be underutilized due to time requirements. While reduced item sets have been studied using Fisher information from Item Response Theory (IRT) models, optimal selection methods remain unclear.

We compared three methods for selecting an optimal subset of items, with the aim of minimizing the uncertainty in the estimates of the disease severity: Ranking by the Fisher information, coordinate descent local search to directly minimize estimate uncertainty, and adaptive selection.

Whereas item ranking based on the expected Fisher information outperformed random choice of items, we saw further gains with the coordinate descent algorithm that directly minimizes the uncertainty of the disease severity estimate. An adaptive algorithm gave an additional slight gain compared to the coordinate descent method. However, the performance of the adaptive method is a best-case limit as we assume that we find the optimal set for the true latent trait scores. For a 5-item subset, the ranked Fisher information method reduced the expected standard deviation by 14 percent compared to random item selection. The corresponding reductions for coordinate descent and adaptive selection were 26 percent and 34 percent respectively.

More sophisticated selection methods substantially improved estimate accuracy for small item sets, with diminishing returns for larger subsets. Because item parameters are retained from the full test, reduced item sets measure the same latent construct as the original test. The choice of method entails a trade-off between methodological complexity and precision.

\end{abstract}


\newpage

\section{Introduction}

In the healthcare system, questionnaires can be important to collect data about the symptoms and experiences of patients. However, questionnaires are also associated with a response burden, i.e., the burden of answering a large number of question items. In addition to the time cost, some participants find it difficult to complete the full battery of questions due to survey fatigue. Survey fatigue is a well-examined phenomenon that disproportionately affects respondents with lower income, lower educational level and lower proficiency in the language of the survey \citep*{le_when_2021}. In this article we look specifically at question items in the context of tracking the progression of Parkinson's disease.

In Parkinson's disease, associated with the loss of dopamine-producing neurons, slowness in movements (bradykinesia) is a core symptom. In addition, Parkinson's manifests through either tremor or rigidity, or both. The disease can also cause non-motor symptoms such as loss of smell and sleep disorders. The disease progresses over time, with tremor as a common early symptom, followed by bradykinesia and rigidity, and later postural instability. A patient with Parkinson's disease may find it difficult to tap their fingers and toes rapidly, and to walk steadily. It may also be a challenge to rise out of a chair. The symptoms may be asymmetric, affecting one side of the body more than the other. For most patients, the disease causes disability within ten years \citep{zafar_parkinson_2025}.

To track the progression of a patient's Parkinson's disease, a unified measure of the symptoms can be obtained using the Unified Parkinson Disease Rating Scale (UPDRS), which was developed in the 1980s and later revised into the Movement Disorder Society Sponsored UPDRS (MDS-UPDRS) \citep{goetz_movement_2008}.

The MDS-UPDRS questionnaire has been estimated to take approximately 30 minutes to complete. Some parts are intended for patients to fill out, and other parts require participation from both the patient and a professional test administrator. It has been indicated that use of the UPDRS test is mostly confined to treatment conducted in a research context \citep{almahadin_parkinsons_2020}. This raises questions about whether a reduced set of test items would enable a more time efficient tracking of symptoms, and thereby facilitate more frequent use of the questionnaire.

Item Response Theory models are frequently employed to evaluate the usefulness of individual items that are a part of a test or a questionnaire. Unlike the classical approach of measuring a latent trait as the sum of item scores, typical IRT models, such as the 2-parameter logistic (2-PL) model, associate each item with parameters that measure the difficulty and discriminatory power of the item. The term \textit{difficulty} comes from the educational application where items measure ability, so that a difficult item is an item that only respondents with high ability are likely to answer correctly. When the latent trait that we measure is the severity of a disease, a more difficult item is instead an item that only patients in a more severe state answer affirmatively. The discriminatory power measures how well the item differentiates between patients with different levels of the latent trait. The estimated values of the item parameters imply an item information function, also called an item characteristic function. This function measures the Fisher information that we obtain about a respondent conditioned on the respondent's latent trait score. More Fisher information implies less uncertainty in the estimate of the latent trait.

The item information is commonly used to decide which items to retain if we want to reduce the item set. \cite{ueckert_improved_2014} concluded from the item information functions that 6 out of 13 items provided 90 percent of the information about the cognitive ability for a studied population. Similarly, \cite{arrington_performance_2020} used the expected item information in a population to rank items that measure motor function disability in Parkinson's patients. Although the item information was more evenly distributed among the 34 items examined in this study, a reduced data set with 65 percent of the items retained 80 percent of the expected Fisher information. These examples demonstrate how IRT models provide valuable information that we cannot extract when we use the summed score as a metric for the latent trait. 

Whereas selecting a subset of items based on the expected Fisher information is a reasonable method, it is not equivalent to selecting the set of items that minimizes the expected uncertainty of the latent trait estimate. This is because the expected Fisher information from an item set is merely the sum of the expected information from individual items. When we add an item to a set, the expected Fisher information of the set will increase the most if we add the item that individually has the highest expected Fisher information, without regard to the properties of the other items in the set. This can lead to a set of items that are all good at measuring in the same region of the latent trait scale, while none of the items are good at measuring in other regions of the scale.

In this article we will explore different methods of assembling an optimal subset of items, with the goal to minimize the expected uncertainty of the latent trait estimates. We do this with MDS-UPDRS data, aiming to select a subset of items that measure motor symptom severity in Parkinson's patient with an uncertainty that is as small as possible.

We compare three methods for selecting an optimal subset of $K$ items from a larger item set. The first method is to rank the expected Fisher information for all items and then select the $K$ most informative items. The second method is to use an algorithm that directly aims to minimize the expected standard deviation of the latent trait estimates with a cyclic coordinate descent algorithm. Both methods assemble subsets of items that are optimized for the population of respondents, for a scenario where all patients are presented with the same subset of items. The third method selects a subset of items based on our prior belief about the latent trait score of an individual patient. Given this prior belief, maximizing the Fisher information is equivalent to minimizing the standard deviation of the post-test estimate. As a benchmark, we also include a scenario with randomly selected items.

In the following sections, we will first outline the test design and describe the setting where the explored methods are applicable. We will then describe the Parkinson's disease data obtained from the Parkinson's Progression Markers Initiative (PPMI). In the Method Preliminaries section, we give a brief introduction to the graded response model which belongs to the IRT family. We also introduce Generalized Linear Mixed Models (GLMM).

In the Methods section, we first establish with a basic example that maximizing the expected Fisher information does not imply that we minimize the expected standard deviation of the latent trait estimates. We then show how the graded response model can be integrated into a GLMM framework to model the change in Parkinson's motor symptoms over time. We describe how we use Item Response Theory and Bayesian inference to estimate this highly parameterized GLMM model in two steps, and look at some model diagnostics to ensure that the fitting process converged in a satisfactory way.

In the results section, we examine the parameter estimates for the items and for the respondents over time. We confirm that the theoretical distinction between maximizing expected Fisher information and minimizing the expected standard deviation of the latent trait estimates is relevant for this data, and then proceed to compare three different methods for selecting a subset of the items.

\section{Test design}

For the methods that we explore, it is assumed that we have access to a dataset from Parkinson's patients who have filled out the full MDS-UPDRS questionnaire. From this data, we estimate item properties and changes in symptom severity over time for patients. As we are interested in finding a subset of the items that allows us to make good estimates for future patients, we are primarily interested in the item properties, and also in symptom progress over time in the population.

When future patients and test administrators fill out a reduced questionnaire, this newly collected data, in combination with established estimates of item parameters and population growth parameters, is used to estimate the individual's current and future motor disability.

\section{Data}

The Parkinson's Progression Markers Initiative (PPMI) provides open-access datasets with information about patients diagnosed with Parkinson's Disease along with healthy controls and non-diagnosed patients with risk factors. The PPMI data include questionnaire data that records the severity of symptoms of Parkinson's disease. The questionnaire uses the Movement Disorder Society Unified Parkinson's Disease Rating Scale (MDS-UPDRS). The datasets used in this article are part of the PPMI Clinical collection, as opposed to the collections PPMI Remote and PPMI Online.

Each question item rates the severity of a symptom or a difficulty caused by a symptom. Each symptom or difficulty is graded on a 5-level scale. For all items that we use in this article, the response options are the same: (0) Normal, (1) Slight, (2) Mild, (3) Moderate and (4) Severe. However, each item provides descriptions of the response options tailored to the specific question.

As an example, Item 2.7 in the questionnaire reads "Over the past week, have you usually had problems with balance and walking?". The first response option is "Normal: Not at all (no problems)", and the last response option is "Severe: I usually use the support of another person to walk safely without falling."

In this article, we follow the example of \cite{arrington_performance_2020} and look at 34 questionnaire items that are all associated with motor function. A subset of these items is found in the dataset MDS-UPDRS Part II Patient Questionnaire: Motor Aspects of Daily Living. The items in this dataset are intended to be completed by the patient and then reviewed for clarity. Another subset of the items is found in the dataset MDS-UPDRS Part III: Treatment Determination and Part III: Motor Examination. These items are administered by a professional who rates the patient's symptoms. The full MDS-UPDRS questionnaire with instructions for test administrators is described by \cite{goetz_movement_2008}.

The two datasets MDS-UPDRS Part II and MDS-UPDRS Part III included in total 4343 patients. From this original set we removed 60 patients who did not meet our requirements. The requirements for remaining in the data were that a patient had completed all items in the two parts of the MDS-UPDRS questionnaire at one or more visits. After removing patients who did not meet this criterion, 4283 patients remained in the data.

Sometimes questionnaire items were answered more than once for a patient in the same day. In those cases, we followed an example from the PPMI data user guide and kept the observation with the highest total score. This represents the worst case during that day, which may be a result of a medication "off" period \citep{parkinsons_progression_markers_initiative_parkinsons_2024}. As a result of the decision to keep only one observation per patient and visit, 231 observations were removed from MDS-UPDRS Part II, and 3849 observations were removed from MDS-UPDRS Part III.

Each of the 4283 patients had at the time of enrollment been assigned to one of four different cohorts: 1521 patients were in the Parkinson's disease cohort, 338 were in the healthy control cohort, and 2343 had risk factors but no Parkinson's diagnosis. A smaller group of 81 patients had a Parkinson's diagnosis, but without dopaminergic deficit on visual inspection. In our analyses, we include all cohorts and do not treat them as separate groups. We do not need to analyze the groups separately, since we will use a longitudinal model with an individual intercept and an individual slope for each patient's development over time. This provides sufficient flexibility to accommodate a variety of patterns for symptom development allowing for different distributions in these groups for the individual intercepts and slopes.

Our final dataset included a total of 755 582 responses, where the lower response options dominated. For the combined item set, about 64 percent of the responses were at level 0 on the 5-level scale. Approximately 23 percent were at level 1, 10 percent at level 2, 2.8 percent at level 3 and 0.3 percent at level 4.

\begin{figure}
\begin{center}
\includegraphics[scale=0.7]{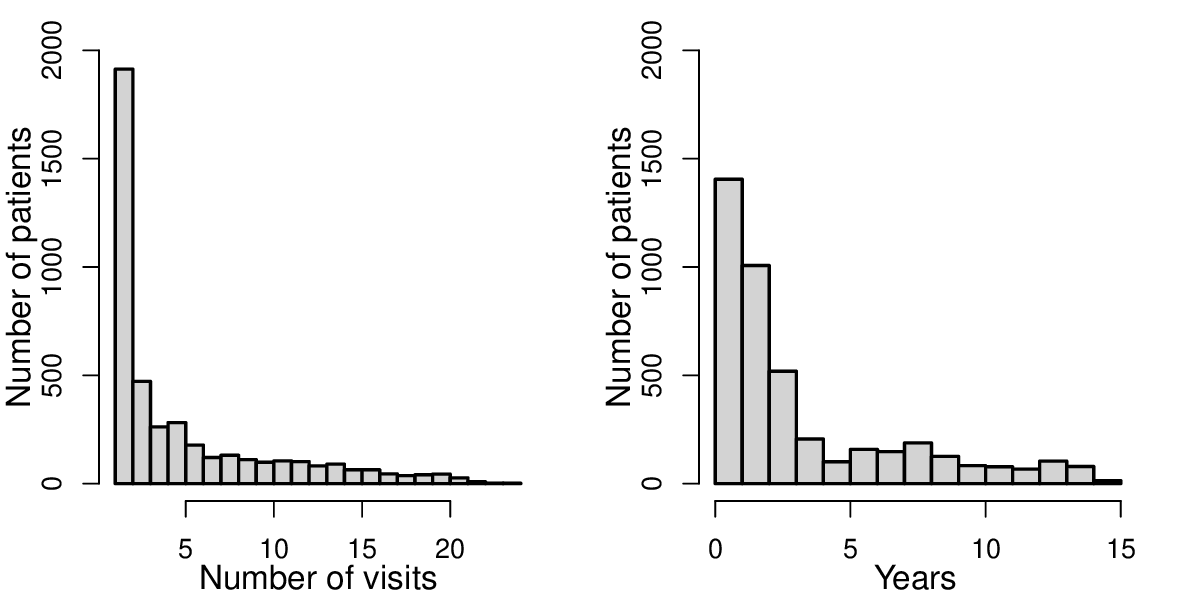}
\caption{Left: The distribution of the total number of visits by the patients in the data. Right: The distribution of the time span between first and last visit included in the data.}
\label{fig:data_visits}
\end{center}
\end{figure}

\section{Method preliminaries} \label{Method preliminaries}

\subsection{The Graded Response Model}

One of the standard IRT models is the 2-PL model, where each item $i$ is associated with a difficulty parameter $b_i$ and a discrimination parameter $a_i$. If we know the latent trait score $\theta_{j}$ of a respondent, we can calculate the probability of a correct answer as
\begin{equation}
\label{eqn:grm1}
\begin{split}
P(y_{i, j} = 1|\theta_j) = \cfrac{e^{a_i(\theta_j - b_{i})}}{1 + e^{a_i(\theta_j - b_{i})}}={\rm logistic}(a_i(\theta_j - b_{i})) .
\end{split}
\end{equation}

This 2-PL model is useful for items where the response can be either correct or incorrect. However, when the responses are measured on a graded scale with levels $0,\dots,M$, an item no longer has one single difficulty parameter value. Instead, each response level is associated with its own difficulty, and the difficulty of reaching level $m$ or above is now
\begin{equation}
\label{eqn:grm2}
\begin{split}
P(y_{i, j} \geq m|\theta_j) = \cfrac{e^{a_i(\theta_j - b_{i,m})}}{1 + e^{a_i(\theta_j - b_{i,m})}} .
\end{split}
\end{equation}

Thus, the formula for reaching level $m$ or any level above is calculated just as we calculate the probability of a correct answer in the 2-PL model. Whereas we speak of the difficulty of passing a single threshold in the 2-PL model, we now speak of the difficulty of passing the threshold for level $m$. An assumption is that when you pass the threshold to reach level $m$, this implies that you also pass the thresholds for all levels below $m$. Likewise, if your response is at level $m$, it implies that you fail to pass the thresholds for all levels above $m$. The probability that your response is exactly at level $m$ in the ordinal scale is
\begin{equation}
\label{eqn:grm3}
\begin{split}
P(y_{i, j} = m|\theta_j) = P(y_{i, j} \geq m|\theta_j) - P(y_{i, j} \geq m +1|\theta_j),
\end{split}
\end{equation}
which is to say that the probability of a response at level $m$ is the probability that you pass the threshold necessary to reach level $m$ but fail to reach the next level, which is level $m+1$. This model is known as the Graded Response Model (GRM) \citep{samejima_estimation_1969}. It assumes that the discrimination parameter is the same for all thresholds that belong to the same item.

The information obtained from an item is a function of the respondent's ability. For the GRM model, the item information conditioned on an ability $\theta_j$ can be calculated with expression 2.48 in \cite{reckase_multidimensional_2009} as
\begin{equation}
\label{eqn:grm_iteminfo}
\begin{split}
I_i(\theta_j) = \sum_{m=1}^{M+1} \cfrac{\left(a_i P_{i, m-1}(\theta_j)(1 - P_{i, m-1}(\theta_j)) - a_i P_{i, m}(\theta_j)(1 - P_{i, m}(\theta_j))\right)^2}{P_{i, m-1}(\theta_j)-P_{i, m}(\theta_j)},
\end{split}
\end{equation}
where $M$ is the number of thresholds and $M+1$ the number of levels in the ordinal scale for item $i$. $P_{i, m}(\theta_j)$ is shorthand for $P(y_{i, j} \geq m|\theta_j)$. It will always be the case that $P_{i, 0} = 1$, since it is trivially easy to reach the lowest level. It is also always the case that $P_{i, M+1} = 0$, since $M$ is the highest level that the respondent can reach. The information obtained from a set of items, conditioned on a latent trait score, is the sum of the information obtained from the individual items. The square root of the inverse of this information is the standard deviation of a latent trait estimate based on the same item set.

\subsection{Generalized Linear Mixed Models for IRT}

Linear Mixed Models (LMM) are a class of linear models that include both fixed effects, that describe a group mean, and random effects that describe how individual subjects deviate from this mean. A common use of these models is to describe change over time. If the response data follow a distribution other than the normal distribution, such as the categorical distribution of an ordinal response variable, we can use a Generalized Linear Mixed Model (GLMM) \citep{mcculloch_chapter_2003}.

A GLMM model for a standard logistic model can be formulated
\begin{equation}
\begin{aligned}
\label{eqn:glmm1}
P(y = 1) = \cfrac{e^{\boldsymbol{x}^\intercal \boldsymbol{\beta} + \boldsymbol{z}^\intercal\boldsymbol{u}}}{1 + e^{\boldsymbol{x}^\intercal \boldsymbol{\beta} + \boldsymbol{z}^\intercal\boldsymbol{u}}}\;\;, 
\boldsymbol{u} \sim {\rm Normal}(\boldsymbol{0}, \boldsymbol{\Sigma})\;\;,
\end{aligned}
\end{equation}
where $\boldsymbol{\beta}$ is a vector of fixed effect parameters, $\boldsymbol{x}^\intercal$ is a vector of fixed effect covariates, $\boldsymbol{u}$ is a vector of random effect parameters, and $\boldsymbol{z}^\intercal$ is a vector of random effect covariates. We assume that the random effects jointly follow a multivariate normal distribution with the mean $\boldsymbol{0}$ and covariance matrix $\boldsymbol{\Sigma}$.

If we want to implement a standard 2-PL IRT model, we modify the GLMM model to become
\begin{equation}
\begin{aligned}
\label{eqn:glmm2}
P(y_{i, j} = 1|\theta_j) = \cfrac{e^{a_i(\boldsymbol{x}_j^\intercal \boldsymbol{\beta} + \boldsymbol{z}_j^\intercal\boldsymbol{u}_j - b_i)}}{1 + e^{a_i(\boldsymbol{x}_j^\intercal \boldsymbol{\beta} + \boldsymbol{z}_j^\intercal\boldsymbol{u_j} - b_i)}}\;\;,
\end{aligned}
\end{equation}
where we have included an item discrimination parameter $a_i$ and an item difficulty parameter $b_i$. If we compare to the standard 2-PL IRT Formula \ref{eqn:grm1}, we see that the latent trait score parameter $\theta_j$ is equivalent to the expression $\boldsymbol{x}_j^\intercal \boldsymbol{\beta} + \boldsymbol{z}_j^\intercal\boldsymbol{u}_j$ in the GLMM model. Thus, the GLMM formulation allows us to treat the latent trait score as a function of the covariates. In our application, the covariate of interest is time. As we have ordinal data, we also need to take this into account in our model. If we substitute our fixed effects and random effects for $\theta$ in formulas \ref{eqn:grm2} and \ref{eqn:grm3}, and add the threshold index $m$ to the difficulty parameter, we get
\begin{equation}
\label{eqn:glmm3}
\begin{split}
P(y_{i, j} = m) =  \cfrac{e^{a_i(\boldsymbol{x}_j^\intercal \boldsymbol{\beta} + \boldsymbol{z}_j^\intercal\boldsymbol{u}_j - b_{i, m})}}{1 + e^{a_i(\boldsymbol{x}_j^\intercal \boldsymbol{\beta} + \boldsymbol{z}_j^\intercal\boldsymbol{u_j} - b_{i, m})}} - \cfrac{e^{a_i(\boldsymbol{x}_j^\intercal \boldsymbol{\beta} + \boldsymbol{z}_j^\intercal\boldsymbol{u}_j - b_{i, m+1})}}{1 + e^{a_i(\boldsymbol{x}_j^\intercal \boldsymbol{\beta} + \boldsymbol{z}_j^\intercal\boldsymbol{u_j} - b_{i, m+1})}}\;\;,
\end{split}
\end{equation}
which is the probability of reaching level $m$ minus the probability of reaching level $m + 1$.

\section{Method} \label{Method}

The MDS-UPDRS questionnaire includes a set of items that assess the severity of the motor disability, which is a latent construct that we use to explain specific motor symptoms. Using a larger number of question items generally gives a smaller uncertainty in the estimate of the latent trait. However, the items don't contribute equally to reducing the uncertainty of the latent trait estimate. From the difficulty thresholds and the discrimination power of each item, we can calculate the Fisher information function as described in Section \ref{Method preliminaries}.

In \cite{arrington_performance_2020}, the population information of an item $j$ is calculated as
\begin{equation}
\label{eqn:item_population_info}
\begin{split}
I_j = \int_{-\infty}^{\infty} p(D_i) I_j(D_i)dD_i,
\end{split}
\end{equation}
where $D_i$ is the latent trait score for respondent $i$, $I_j(D_i)$ is the item's Fisher information function and $p(D_i)$ is the density function for the latent trait in the population. We will here refer to this as the expected Fisher information of an item, as it is the expectation of the Fisher information obtained from the item with respect to a randomly selected respondent from the population.

If we want to reduce the response burden by presenting a smaller number of items, we want to do this in a way that minimizes the expected uncertainty of the latent trait estimate given the number of retained items. We can measure this expected uncertainty as the expected standard deviation in the estimated latent trait score of a randomly chosen respondent.

One strategy to find an item set of size $K$ that gives a small expected uncertainty is to rank the items by their expected Fisher information, and to select the $K$ items with the highest expected information. However, maximizing the expected Fisher information does not necessarily minimize the expected standard deviation, as shown in the following example.

\subsection{Expected Information and Expected Standard Deviation}

\begin{figure}
\begin{center}
\includegraphics[scale=0.4]{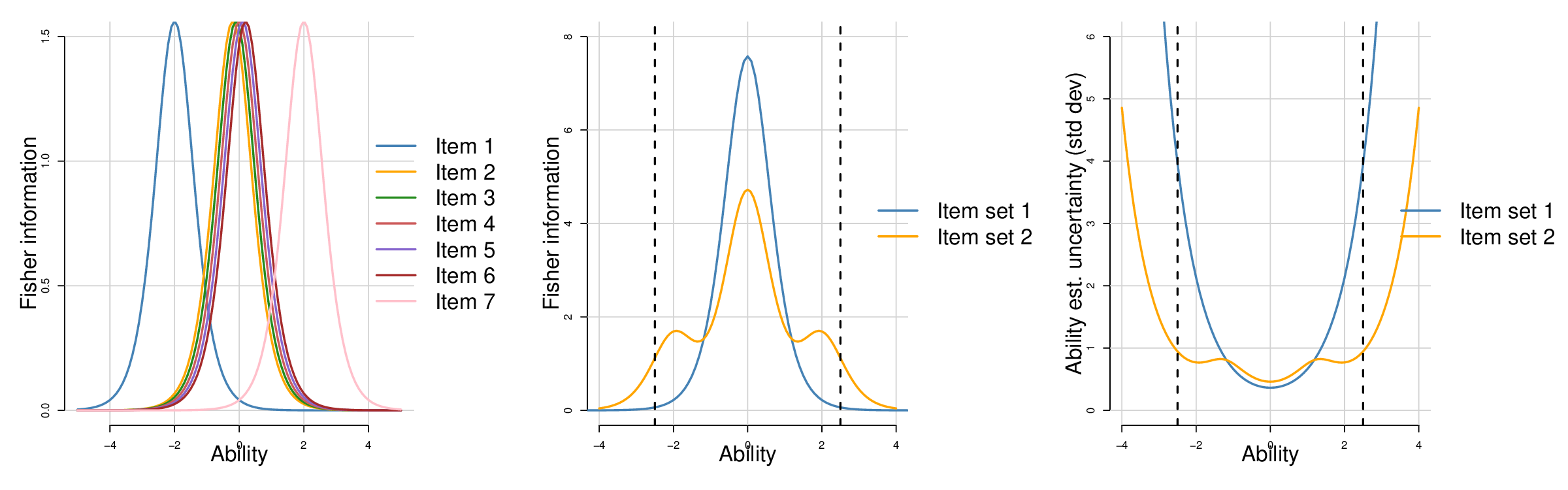}
\caption{Left: The information functions for 7 items in a 2-PL IRT model. Middle: The information function for the set of items 2, 3, 4, 5, 6 (Item set 1), and for the set of items 1, 3, 4, 5, 7 (Item set 2). Right: The standard deviation of the latent trait score estimate as a function of the true latent trait score for the set of items 2, 3, 4, 5, 6 (Item set 1), and for the set of items 1, 3, 4, 5, 7 (Item set 2).}
\label{fig:theoretical_example}
\end{center}
\end{figure}

We will here present a theoretical example with a 2-parameter IRT model and a set of 7 items. As illustrated in Figure \ref{fig:theoretical_example}, 5 of the items have difficulties clustered around 0, with difficulty values -0.2, -0.1, 0, 0.1, 0.2. One easier item has the difficulty -2, and one more difficult item has the difficulty 2. For simplicity, we let all items have equal discrimination power with the value of the discrimination parameter set to 2.5. These item parameters are not set to reflect a specific real scenario, but merely to illustrate a theoretical concept.

Our goal is to select a subset of 5 items that gives a low expected standard deviation for the population of respondents. We assume that the respondents have latent trait scores that follow a standard normal distribution.

The middle chart in Figure \ref{fig:theoretical_example} shows the information function for two different item sets. Item set 1 includes the 5 items with information functions that peak around 0, i.e., items 2, 3, 4, 5, and 6. We see that the information function for item set 1 provides plenty of information about respondents with ability around 0, while it provides almost no information about respondents with ability scores at -2.5 or 2.5 as marked in the figure. In comparison, the information for item set 2, which includes items 1, 3, 4, 5, and 7, is considerably lower around 0, but higher in the tails of the distribution of the latent trait scores.

Assuming abilities that follow a standard normal distribution in the population, we can use formula \ref{eqn:item_population_info} to show that the expected information from item set 1 is 4.07 while the expected information from item set 2 is 2.87. If our optimization method aims to maximize the expected information, we should therefore prefer item set 1 over item set 2.

The right chart in Figure \ref{fig:theoretical_example} shows the standard deviation of the estimated latent trait score as a function of the true ability. We see that item set 1 gives a lower uncertainty for abilities around 0. However, the difference is only slight compared to item set 2, due to diminishing usefulness of additional Fisher information. In the tails, item set 2 gives considerably better precision since items 1 and 7 contribute information at locations where it translates into a substantially reduced standard deviation. Even though we have a latent trait score distribution that gives more weight to scores close to 0, the expected standard deviation for item set 2 (0.64) is lower than that of item set 1 (0.77).

An argument for maximizing the expected Fisher information is the simplicity of this method. The problem of identifying the optimal set of $K$ items with regard to the population Fisher information is trivial, as the best set will consist of the $K$ items that rank highest individually. That is, the perceived usefulness of an item will be independent of which other items are included in the set. This is no longer the case when we want to assemble the best set of $K$ items to minimize the expected standard deviation of the latent trait estimate.

With a small item pool and a small $K$, a brute-force algorithm that evaluates each possible item set of $K$ items may be feasible. For larger item pools and larger values of $K$, we may have to resort to optimization algorithms that are reasonably fast at finding a local optimum, but with no guarantee of finding the global optimum. By initializing the search with the item set that maximizes the expected Fisher information, we can at least guarantee that we will end up in a position equal to or better than this.

\subsection{Our longitudinal model}

Formula \ref{eqn:glmm3} gives the general form of the GLMM equation to calculate the probability of a response at a specific level that we call $m$. We are now interested in the probability of an observed response $y_{i, j, t}=m \in \{0,\dots,M\}$ to item $i$ by subject $j$ at time $t=0,1,\dots$ (where the item $i=i(j,t)$ is determined by an assignment rule which assigns a specific item to subject $j$ at time $t$). Moving from the general formulation of a GLMM model to terms that are specific for our purposes gives
\begin{equation}
\label{eqn:glmm4}
\begin{split}
P(Y_{i, j, t} = y_{i, j, t}|\boldsymbol{\tau}, \boldsymbol{\gamma}) =  \cfrac{e^{a_i\left(\beta_0 + u_{0, j} + (\beta_1 + u_{1, j}) t - b_{i, y_{i, j, t}}\right)}}{1 + e^{a_i\left(\beta_0 + u_{0, j} + (\beta_1 + u_{1, j}) t - b_{i, y_{i, j, t}}\right)}} -\cfrac{e^{a_i\left(\beta_0 + u_{0, j} + (\beta_1 + u_{1, j}) t - b_{i, y_{i, j, t}+1}\right)}}{1 + e^{a_i\left(\beta_0 + u_{0, j} + (\beta_1 + u_{1, j}) t - b_{i, y_{i, j, t}+1}\right)}}\;\;.
\end{split}
\end{equation}

In this formula, the latent score for subject $j$ is calculated as the sum of two terms. The first term is an individual intercept, that is the sum of the global intercept $\beta_0$ and the individual deviation $u_{0, j}$. In the second term we have an individual slope, that is the sum of the global slope $\beta_1$ and the individual deviation $u_{1, j}$. The individual slope is multiplied by the time variable $t$. The probability is explicitly conditioned on the fixed effect parameters $\boldsymbol{\tau}$ and the random effect parameters $\boldsymbol{\gamma}$. Like the model without treatment effect specified in \cite{arrington_performance_2020}, this is a GRM model where the latent trait $\theta_j$ is defined with an individual intercept and an individual slope that represents the change over time for respondent $j$. 

\subsection{Parameter estimation}
To fit all model parameters jointly, we find the parameter estimates that maximize the expression 
\begin{equation}
\label{eqn:glmm_likelihood}
\begin{split}
 \prod_{j, t, i=i(j,t)} & P(Y_{i, j, t} = y_{i, j, t}, \boldsymbol{\tau}|\boldsymbol{\gamma}) f(\boldsymbol{\gamma})=  \\ \prod_{j, t, i=i(j,t)} & \left(\cfrac{e^{a_i\left(\beta_0 + u_{0, j} + (\beta_1 + u_{1, j}) t - b_{i, y_{i,j,t}}\right)}}{1 + e^{a_i\left(\beta_0 + u_{0, j} + (\beta_1 + u_{1, j}) t - b_{i, y_{i,j,t}}\right)}} -\cfrac{e^{a_i\left(\beta_0 + u_{0, j} + (\beta_1 + u_{1, j}) t - b_{i, y_{i,j,t}+1}\right)}}{1 + e^{a_i\left(\beta_0 + u_{0, j} + (\beta_1 + u_{1, j}) t - b_{i, y_{i,j,t}+1}\right)}}\right) \\ & \cdot (2 \pi)^{-1} \det{(\boldsymbol{\Sigma})}^{-1/2} \exp\left(-(\boldsymbol{u}_j^\intercal \boldsymbol{\Sigma}^{-1} \boldsymbol{u}_j)/2\right),
\end{split}
\end{equation}
where $f(\boldsymbol{\gamma})$ is the density of the random effects $\boldsymbol{\gamma}$ and
\begin{equation}
\label{eqn:glmm_likelihood_random}
\begin{split}
 \boldsymbol{u}_j = 
 \begin{bmatrix}
     u_{0, j} \\
     u_{1, j}
 \end{bmatrix} ,\;\; 
\boldsymbol{\Sigma} =
 \begin{bmatrix}
    \sigma_{u_0}^2 & \sigma_{u_0, u_1} \\
    \sigma_{u_0, u_1} & \sigma_{u_1}^2 
\end{bmatrix}
 \;\;.
\end{split}
\end{equation}

The GLMM model was fitted with MCMC sampling in Stan. Stan estimated the parameters jointly, including the individual random effects. The decision to include all 4283 patients in our dataset therefore generated a total of 8739 freely estimated model parameters as listed in Table \ref{table:parameters}.

Models with a large number of parameters can be prone to identifiability problems. To facilitate a stable estimation process and to minimize identifiability problems, we executed the estimation in two steps. In the first step, which we call a pre-estimation step, we fitted some of the item parameters using a subset of the data that included item responses only from the first visit. In the second step, where Stan estimated the full GLMM model, these pre-estimated parameter values were treated as fixed. By treating some item parameters as fixed in the final model, we provided constraints to facilitate the estimates of the freely estimated parameters.

To further facilitate the estimation process, we used the following priors: The fixed effect for the slope had a truncated normal prior with the mean 0 and the variance 1. The distribution was truncated to the interval [-1, 1]. The variance of the random slope over patients had a truncated normal prior with the mean 0 and the variance 1. The distribution was truncated to the interval [0.01, 2]. In the Stan-specification of the model, the freely estimated item difficulties for levels 3 and 4 were specified as how much larger they were compared to the previous level, e.g., the estimated parameter for level 3 of item $i$ specified how much more difficult this level was to reach compared to level 2 of item $i$. This ensured that difficulty was strictly increasing with higher levels on the ordinal scale. These increases for levels 3 and 4 for all items had a truncated normal prior with mean 1 and variance 1. The distributions were truncated to the interval [0.01, 10]. All priors were set to be weak, as the purpose was to not incorporate prior beliefs, but merely to facilitate convergence in the parameter fitting process.

The large number of parameters made the model slow to fit. However, in a use scenario, we would expect that the full model only needs to be fitted once, or perhaps on a few additional occasions to incorporate new data in the calculation of the group-level parameters. The more frequent process of obtaining an estimate of the symptom severity for a new or returning patient only requires a trivial amount of time. It can be done with a method where the fitted model is used as a prior that we combine with the new data \citep{casella_introduction_1985}.
\begin{center}
\begin{table}
\footnotesize
\begin{tabular}{ |llllll| } 
 \hline
  Model parameters & N  & Fixed & Estimated & Estimated stage 1 & Estimated stage 2 \\  
   \hline
  Mean intercept & 1 & 1 & 0 & 0 & 0 \\
  Mean slope & 1 & 0 & 1 & 0 & 1 \\
  Random intercept variance & 1 & 1 & 0 & 0 & 0 \\
  Random slope variance & 1 & 0 & 1 & 0 & 1 \\
  Random effects correlation & 1 & 0 & 1 & 0 & 1\\
  Item thresholds & 136 & 0 & 136 & 68 & 68  \\
  Item discrimination & 34 & 0 & 34 & 34 & 0  \\
  Individual random intercepts & 4283 & 0 & 4283 & 0 & 4283 \\
  Individual random slopes & 4283 & 0 & 4283 & 0 & 4283 \\
   \hline
  Total & 8741 & 2 & 8739 & 102 & 8637  \\
   
 \hline
\end{tabular}
\caption{Model parameters in the longitudinal mixed effects model used to estimate the trajectories of symptom severities.}
\label{table:parameters}
\end{table}
\end{center}
\subsubsection{Stage 1: IRT estimation from first-visit data}

In a first step, we used only data from the first visit of each patient. From this data, we estimated the item parameters. These estimates were conditioned on the assumption that the patient symptom severities at the initial visit followed a standard normal distribution. This is a standard assumption in IRT parameter estimation when we measure a latent trait at a single time point.

The pre-estimation step was performed with the ltm package in R \citep{rizopoulos_ltm_2007}. This pre-estimation step provided both item parameters and individual symptom severity estimates at the time of the first visit.

Each item has 4 difficulty thresholds and one discrimination parameter. We decided to use the pre-estimates for the two lower thresholds of each item, and also the pre-estimates of the discrimination parameter, as fixed values in the stage 2 GLMM model. The IRT estimation from the first visits generated estimates for thresholds 3 and 4 as well. However, we opted not to use these as the frequency of patients with severe symptoms was low at the first visit. Individual measures of symptom severity at the first visit were also generated for each patient in the pre-estimation step. Nor were these estimates treated as fixed in the final model. We decided to let estimates of symptom severity at visit 1 be freely estimated in the final GLMM model, where the estimates could incorporate information obtained from item responses at later time points.

Although we abstained from using some of the estimates from the stage 1 model, these estimates  still facilitated the stage 2 GLMM estimation by serving as initial values for the MCMC sampling in Stan.

\subsubsection{Stage 2: Generalized Linear Mixed Model}

The GLMM model was fitted with the sample method in Stan (CmdStan v. 2.36.0), accessed through R (v. 4.4.3) and the interface CmdStanR (v. 0.9.0) \citep{gabry_cmdstanr_2025}. We set the Stan MCMC algorithm to run 4 chains, which is the minimum number of chains recommended by \cite{vehtari_rank-normalization_2021}. The sampling algorithm ran for a total of 8000 iterations, of which the first 4000 ran in the warmup phase and were thus discarded from the collection of sample points included in the estimates.

After fitting the GLMM model in Stan, we looked at the R-hat statistic and the effective sample size (ESS). The R-hat statistic, which measures how well the chains converge, was on average 1.0003. The highest parameter estimate had an R-hat of 1.009, which is below the upper threshold of 1.01 suggested by \cite{vehtari_rank-normalization_2021}. The ESS bulk metric was on average 29 103, and the lowest value 758. The ESS tail metric was on average 11 678, and the lowest value 1 764.

\subsection{Item set optimization}

With item parameters and individual ability parameters estimated, we could define the information function for each item. For each item, we could also calculate the expected Fisher information, defined as the expectation of the Fisher information with regard to a randomly selected latent trait estimate.

\subsubsection{The maximum expected Fisher information criterion}

We first selected $K \in \{1, 2, ..., 34\}$ items using the maximum Fisher information criterion. This entailed ranking the items from that with the highest expected Fisher information to that with the lowest. For each value of $K$, we selected a set that included the $K$ top ranked items.

\subsubsection{The minimum expected standard deviation criterion}

Finding the set of $K$ items that gives the lowest expected standard deviation of the latent trait estimate is less straightforward than maximizing the expected Fisher information, as the usefulness of an item depends on the other items in the set. Each potential combination of items must therefore be evaluated as a set and we have a combinatorial optimization problem \citep[Chapter 3]{givens2012computational}. We use an item selection method that we refer to as coordinate descent local search. The method uses the following algorithm:

\begin{enumerate}
    \item Initiate a vector of size $K$. Select $K$ arbitrary items, each of which is inserted as an element in the vector.
    \item Go through all $K$ vector elements. For each position $k$ in the vector, check which item – either the current item in $k$ or one of the unused items – produces the item set with the lowest standard deviation of the ability estimates. Keep that item in $k$.
    \item After going through all positions in the vector, repeat the full procedure. Do this until no further substitution produces a lower expected standard deviation.
\end{enumerate}

If a new item is found in Step 2 which is better at position $k$, an exchange of two items is done. Therefore, this algorithm is a Fedorov-type exchange algorithm \citep{fedorov1972theory,nguyen1992review}. 
Below, the algorithm is described in pseudocode.\\

\begin{singlespace}
\begin{algorithm}[H]
\caption{Implementation of coordinate descent local search}
\label{coordinate_descent}
\SetAlgoLined
\KwIn{
\begin{itemize}
\item We have a pool of $N$ items with estimated difficulty thresholds and discrimination parameters.
\item We have an estimated population ability distribution.
\item $K$ is the number of items in the subset to optimize.
\item \textit{subset} is a vector of $K$ elements, where each element represents an item.
\end{itemize}
}
\KwResult{
\begin{itemize}
\item A locally optimal subset of $K$ items, where the optimal subset is one that minimizes the expected standard deviation of the ability estimate from a random respondent.
\end{itemize}
}

\BlankLine

\textit{Step 1: Randomly select $K$ items and store them in a vector}

\BlankLine

$i \leftarrow 0$\;
$\text{subset}_i \leftarrow \text{select.random}(\text{n.select}=K, \text{from}=1:N)$\;

\BlankLine

\textit{Step 2: Substitute items in the vector $\text{subset}$ until a local optimum is reached.}

\BlankLine

\Repeat{$\text{subset}_i[k] = \text{subset}_{i-1}[k] \quad \forall k$}{
    $i \leftarrow i + 1$\;
    $\text{subset}_i \leftarrow \text{subset}_{i-1}$\;
    \For{$k \in \{1, 2, ..., K\}$}{
        $\text{best.sd} \leftarrow \text{calculate.expected.sd}(\text{subset}_i)$\;
        $\text{best.item} \leftarrow \text{subset}_i[k]$\;
        \For{$n \in \{1, 2, ..., N\}$}{
            \If{$n \notin \text{subset}_{i}$}{
                $\text{subset}_i[k] \leftarrow n$\;
                $\text{this.sd} \leftarrow \text{calculate.expected.sd}(\text{subset}_i)$\;
                \If{$\text{this.sd} < \text{best.sd}$}{
                    $\text{best.item} \leftarrow n$\;
                    $\text{best.sd} \leftarrow \text{this.sd}$\;
                }
            }    
        }
        $\text{subset}_i[k] \leftarrow best.item$\;
    
    }
}

\Return{$\text{subset}_i$}\;

\BlankLine
\end{algorithm}
\end{singlespace}

\BlankLine

Algorithm \ref{coordinate_descent} is not guaranteed to find the global optimum. However, it will find a local optimum, and by running it multiple times with different initial item subsets we can make sure that it ends up with the same result each time.

\subsubsection{Adaptive design}

We also explored an adaptive approach where each patient gets a tailored item subset, which gives the most information conditioned on estimated symptom severity. The results for the adaptive design assume that the optimal item set chosen based on the current latent trait estimate is the same as if it has been chosen based on the true latent trait score. Therefore, the standard deviations obtained from the adaptive design reflect a best-case scenario. The standard deviations represent the limit of what an adaptive algorithm can achieve.

\section{Results} \label{Results}

\subsection{Parameter estimates}

\begin{figure}
\begin{center}
\includegraphics[scale=0.7]{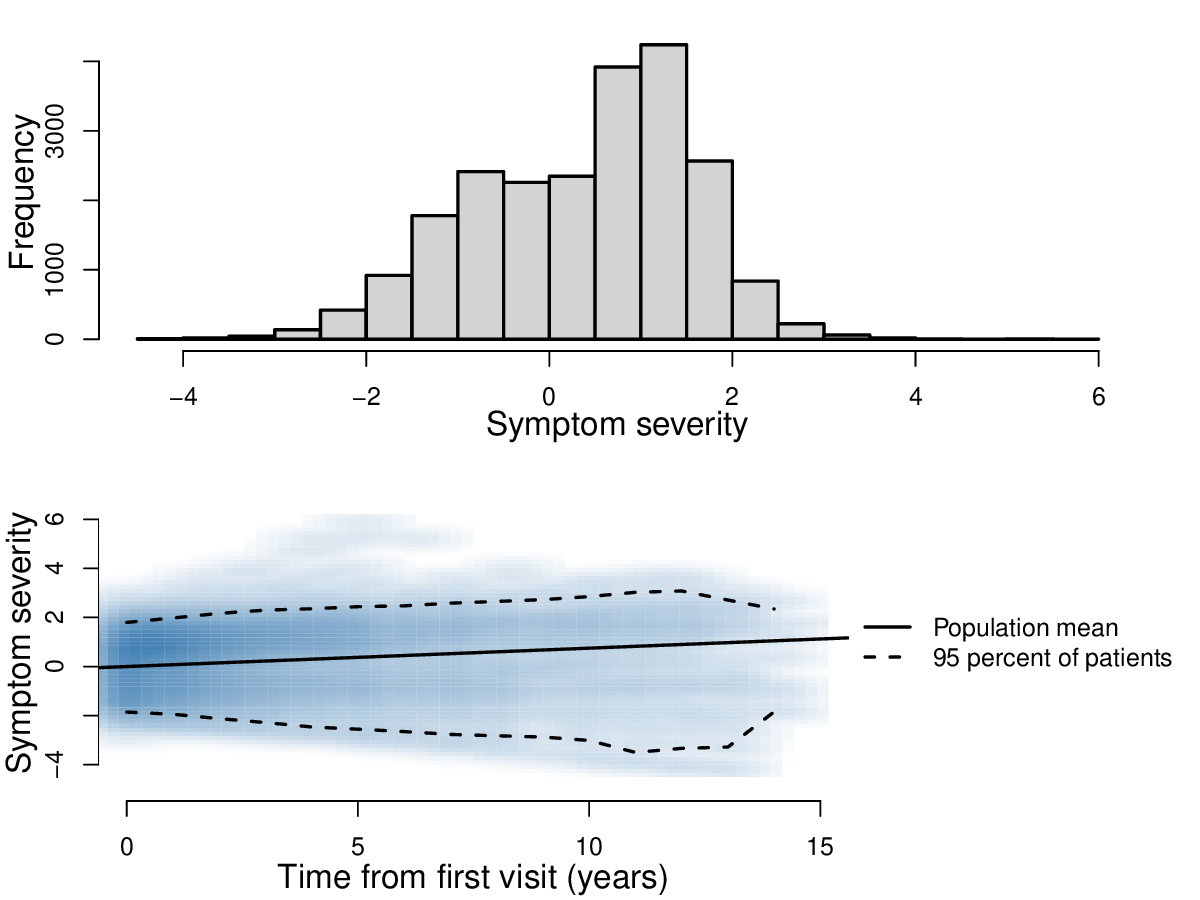}
\caption{Top: The estimated distribution of symptom severity in the population. All measurement time points are included. Bottom: The estimated distribution of symptom severity in the population over time. Darker shades represent a higher density of estimates.}
\label{fig:theta_distribution}
\end{center}
\end{figure}

The histogram in Figure \ref{fig:theta_distribution} shows the distribution of motor symptom severities, where each observation is one patient at one time point. The mean estimated severity is 0.38. A mean above 0 was expected since the model assumed that the mean was 0 at the first visit, and on average symptoms got worse over time for patients with Parkinson's disease.

The fitted model gave us estimates of symptom severities in a range from $-4.2$ to $5.9$. These symptom severities were measured on a scale standardized so that the population of respondents at their first visit followed a standard normal distribution.

Table \ref{table:glmm_estimates} shows the fixed and random effects estimated for the GLMM model. Regarding the population mean slope, which was estimated as positive 0.075, we should recall that this average is estimated from both patients with Parkinson's disease and healthy controls. It does not represent the expected progression of motor symptoms for patients who have been diagnosed with Parkinson's disease. The mean intercept and mean slope for each cohort, after including individual random effects, are also presented in the table.

The bottom chart in Figure \ref{fig:theta_distribution} shows how the estimated symptom severities changed over time. We may note in the figure that the patients with the fastest symptom progression typically leave the study in less than 7 years.

\begin{figure}
\begin{center}
\includegraphics[scale=0.3]{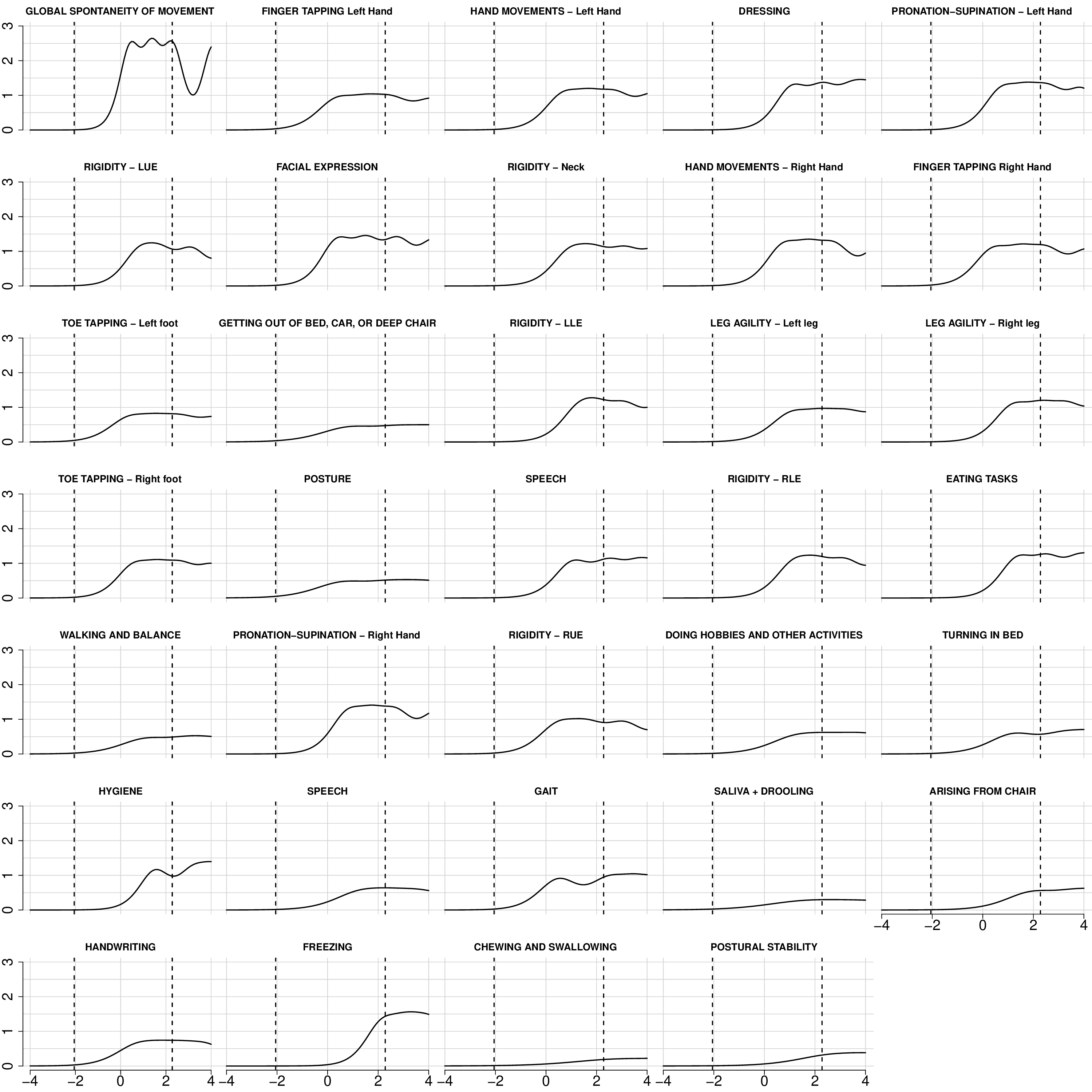}
\caption{Each plot shows the Fisher information of an item as a function of motor symptom severity. The dashed vertical lines capture 95 percent of the symptom severity estimates.}
\label{fig:item_information}
\end{center}
\end{figure}

Figure \ref{fig:item_information} displays the estimated Fisher information for each item as a function of the motor symptom severity. The items are presented in the same order as in \cite{arrington_performance_2020}, Figure 1. When we compare our information functions with those of Arrington et al., we see a similar overall pattern. However, there are also notable differences. One difference is that the item \textit{Global spontaneity of movement} stands out in our estimate as exceptionally informative in the upper central range of abilities. A systematic difference is that \cite{arrington_performance_2020} has several items that measure low levels of motor symptoms reasonably well. In our estimates we do not see any such items, as we have no items with substantial Fisher information to the left of the 95 percent lower bound for the symptom severity distribution. This is likely a result of different patient inclusion criteria. Here we use data from each visit that generates a complete questionnaire, and we include patients without a Parkinson's disease diagnosis. The analysis in \cite{arrington_performance_2020} only included 423 newly diagnosed Parkinson's patients. These patients had been on PD medication for 60 days or less before the baseline measurement.

\begin{center}
\begin{table}
\footnotesize
\begin{tabular*}{\textwidth}{@{\extracolsep{\fill}} |lll|lll| }  
 \hline
  Fixed effects & Estimate & Std dev & Random effects & Estimate & Std dev \\  
   \hline
  Fixed intercept & 0* & & Random Intercept & 1* &  \\
  Fixed slope & 0.075 &  0.004 & Random slope & 0.027 & 0.001 \\
  & & & Correlation & 0.085 & 0.024 \\
 \hline
\end{tabular*}

\begin{tabular*}{\textwidth}{@{\extracolsep{\fill}}|llllll|} 
    \hline
  Cohort means & PD diagnosis & PD risk factors & Healthy controls & & \\  
   \hline
  N & 1521 & 2343 & 338 & &  \\
  Mean intercept & 1.001 & -0.505 & -1.166 & &  \\
  Mean slope & 0.100 &  0.065 & 0.0373 & & \\
 \hline
\end{tabular*}

\caption{Parameter estimates from the longitudinal GLMM model. Values marked with an asterisk are fixed rather than freely estimated. The cohort means are the sums of the fixed effects and the mean individual random effects within each cohort. A group of 81 patients, described in the section \textit{Data}, does not belong to any of the three cohorts included in the table.}
\label{table:glmm_estimates}
\end{table}
\end{center}

\subsection{Optimal item sets}

We optimized the item set both with regard to the expected item information and with regard to the expected standard deviation. The two different optimization criteria produce different item sets. This confirms that ranking the items based on expected Fisher information does not provide item sets that give the lowest possible uncertainty in the latent trait estimates.

\begin{figure}
\begin{center}
\includegraphics[scale=0.3]{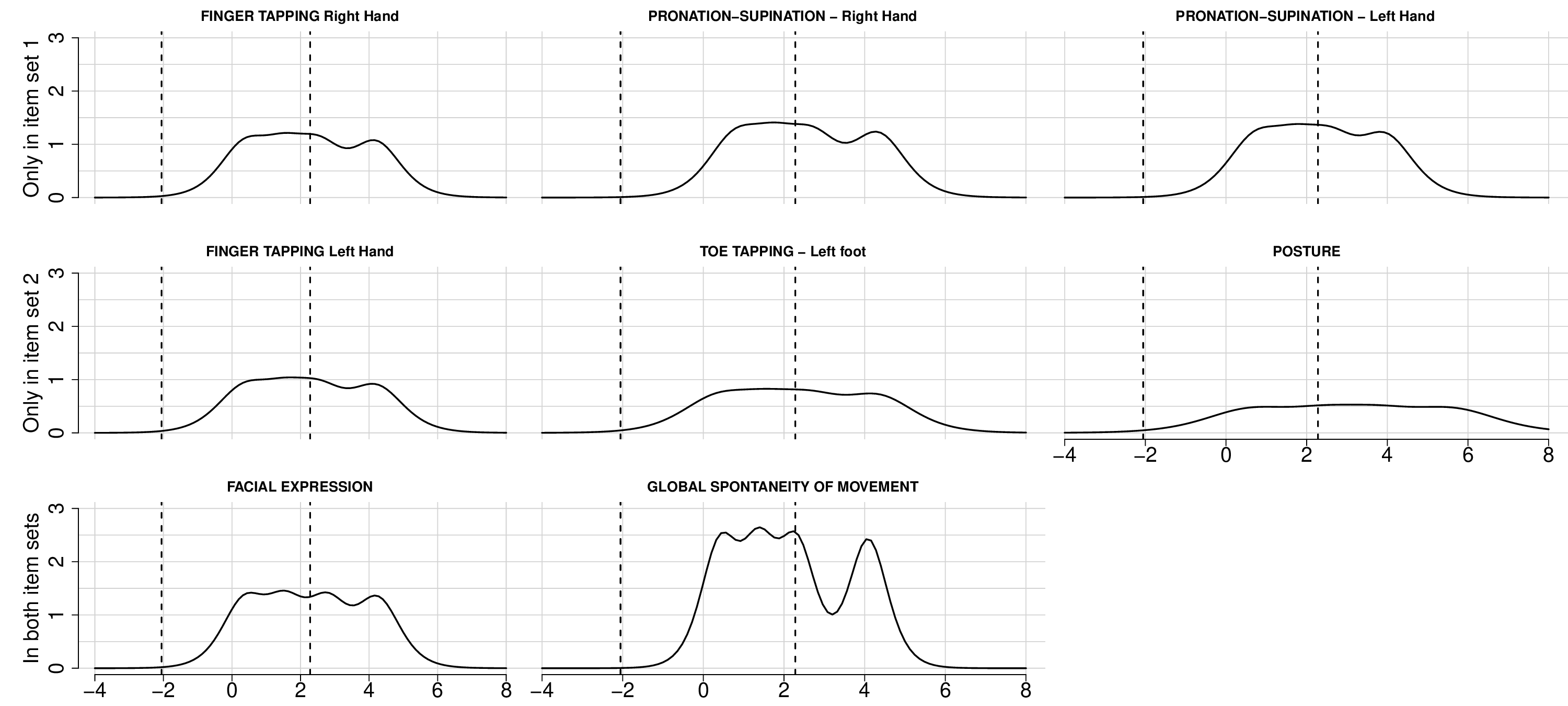}
\caption{Items included in an optimal set of size 5 with different optimization methods. The items on the top row in the figure are included only when we select items to
maximize the expected Fisher information, and the items on the second row are included only when we select items to minimize the expected standard deviation. The two items on the third row are included in the set of 5
items for both methods.}
\label{fig:select5}
\end{center}
\end{figure}

Figure \ref{fig:select5} illustrates how the choice of items differs depending on the selection method for the case where we choose 5 of the 34 items. The items on the top row in the figure are included only when we select items to maximize the expected Fisher information, and the items on the second row are included only when we select items to minimize the expected standard deviation. The two items on the third row are included in the set of 5 items for both methods. If we compare the top two rows, we see that items selected to minimize the standard deviation for the latent trait estimates tend to have a flatter information function. The clearest example is the item that rates the posture of the patient. Whereas this information function has a lower peak than the other items in the figure, it spreads over a wider interval. Therefore, it provides more information in the tails, where most other items provide little or no information.

\begin{figure}
\begin{center}
\includegraphics[scale=0.7]{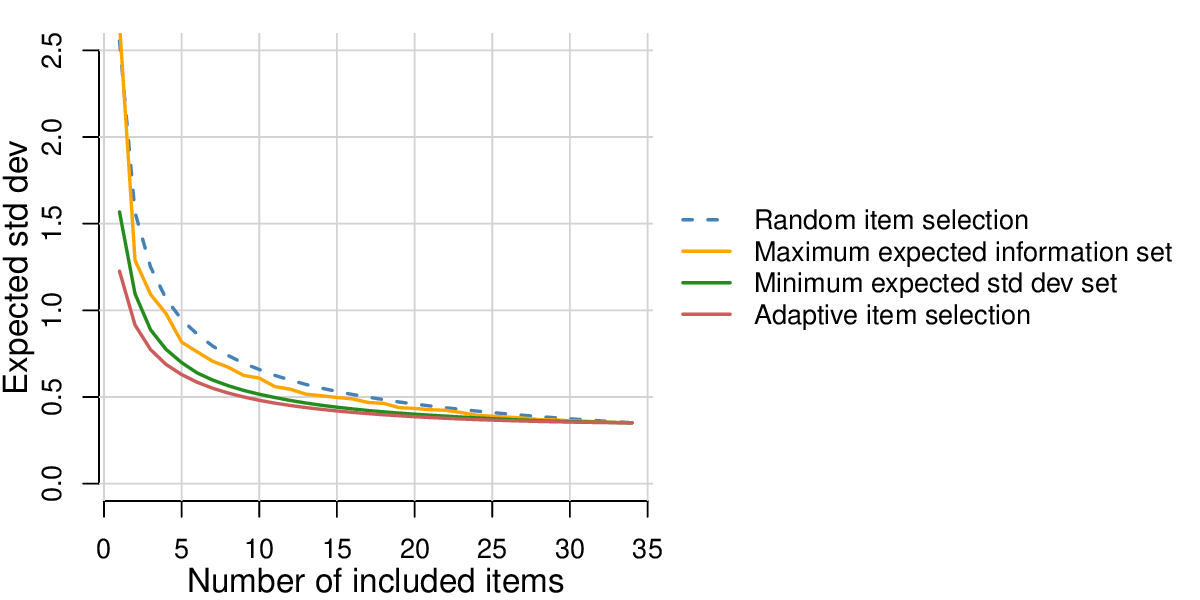}
\caption{The expected standard deviation of an ability estimate with different methods for assembling optimal item sets for a specific size.}
\label{fig:compare_methods}
\end{center}
\end{figure}

Figure \ref{fig:compare_methods} shows how the expected standard deviation decreases as we include additional items in the item set. Each of the four lines in the graph represents one method of assembling an optimal set of $K$ items where $K = 1, 2, ..., 34$. One of the lines represents a random selection of items. The expected standard deviation for random item selection was obtained by adding items in random order and calculating the expected standard deviation for each cumulative item set. This procedure was repeated 200 times, and the result presented in Figure \ref{fig:compare_methods} is the average over these 200 repetitions. The expected standard deviation obtained through random item selection is included as a benchmark.

Compared to random selection, we make a slight gain in accuracy by ranking the items by expected Fisher information and including the $K$ most informative items in each set. The gain is larger if we instead assemble the items by minimizing the expected standard deviation through a coordinate descent algorithm.

We get a further slight improvement if we use an adaptive item selection algorithm. The adaptive algorithm selects a set of items for a respondent at a time point based on the current latent trait estimate. Here, the adaptive algorithm makes the item selection based on latent trait estimates obtained from the full data. As the expected standard deviations from the adaptive algorithm represent a best case, we should view it as the limit of what may be achieved by adaptive item selection. With its modest improvement over the static approach, it is unclear whether the adaptive algorithm would in practical application give better precision.

A practical implication of these results is that if we accept an increase in the standard deviation of the estimate from 0.35 to 0.46, we could cut the number of items substantially. With random item selection, the test would require 20 items instead of 34. If we instead select items based on the maximum information criterion, 19 items are required for the same precision. With the minimum standard deviation criterion and the coordinate descent algorithm, we are required to use 14 items, which means that we only need about 40 percent of the full test item battery. With the adaptive algorithm we could in a best case scenario do even better, although marginally. 

The advantage of optimizing the item set, and the use of more complex methods, dissipates as the number of included items increases. Table \ref{table:optimal_set_comparison} shows that with only 5 items included, maximizing the expected Fisher information decreases the expected standard deviation by 14 percent compared to random item selection. The corresponding decreases of coordinate descent and adaptive selection were 26 and 34 percent respectively. With 10 items in the set, the corresponding decreases were 8, 22 and 27 percent. With 20 items, the decreases were 6, 13 and 16 percent compared to random item selection.

\begin{center}
\begin{table}
\footnotesize
\begin{tabular*}{\textwidth}{@{\extracolsep{\fill}}|llll|} 
  \hline
  & \multicolumn{3}{c|}{Decrease in expected standard deviation, percent} \\
\cline{2-4}
  \# items & Max Fisher info & Coordinate descent & Adaptive selection \\  
   \hline
  5 & 14 & 26 & 34  \\
  10 & 8 & 22 & 27 \\
  20 & 6 &  13 & 16 \\
 \hline
\end{tabular*}

\caption{The percentage decrease in expected standard deviation of the latent trait scores, compared to the expected standard deviation obtained from random selection of items.}
\label{table:optimal_set_comparison}
\end{table}
\end{center}

\section{Discussion}

The choice of items in a questionnaire is crucial if we want to reduce the response burden while minimizing the loss of precision in the estimate. The item response function is an excellent tool for making the item selection. We have explored here three different methods to select a good subset of items from an item pool, and whereas we can say a methodical selection is better than a random selection of items, we cannot say that one of the explored methods is superior. Instead, the choice of method is a trade-off between accuracy and simplicity. We have consistently used the expected standard deviation of the latent trait estimates of the respondents as the measure of model accuracy. This ensures that we give consideration also to respondents in the tails of the distribution, unlike the expected Fisher information, which favors items that are informative in the center of the distribution of the latent trait.

In our results, the adaptive method of item selection gave slightly better accuracy than an optimized static test, but this was based on the assumption that we could assemble the item set that is optimal for the true latent trait score. In a real scenario, the items are optimized for a provisional estimate of the latent trait score. We cannot know whether this is the optimal set for the true latent trait score. Therefore, the results for the adaptive approach shows the limit of what the method can achieve. The possible gains come at the price of a more elaborate routine, where the item selection process must be repeated for each time that the questionnaire is to be filled out. In a real application, the item selection would need to be automated, which makes it better suited for digital questionnaires than for the pen-and-paper variety.

A good alternative to the adaptive method is optimizing the item set with regard to the expected standard deviation with a local search algorithm. The accuracy of this option was only slightly below the performance of the adaptive method. In practical use cases, it provides more simplicity. Optimizing a set of items of a certain size for the entire population requires a non-trivial optimization algorithm. However, once the item set is assembled, the same item set can be used for all patients at all time points. One can also imagine a hybrid solution with an element of adaptive selection, where two or more item sets are assembled. In the case of two item sets, one could be optimized for patients with milder symptoms and one for patients with severe symptoms, and the choice between these tests can be made based on previous test scores or based on a clinical diagnosis.

Assembling a reduced item set that maximizes the expected Fisher information further simplifies the selection process compared to local search optimization with regard to the standard deviation. The additional simplicity comes at the price of greater expected uncertainty in the latent trait estimates, which will be seen in the tails of the distribution of the latent trait. Although not as good as more complex methods, the maximization of expected Fisher information still produces better estimates than a random choice of items. It can be a good option when simplicity is valued, for example if the item pool is frequently updated, requiring subsequent updates of the item sets.

Regarding the full set of MDS-UPDRS items that measure motor function disability, none of the items are good at measuring mild symptoms. The current items measure symptoms well in the interval from 0 to 5, when using a scale where 0 is the mean level of symptom severity at first measurement occasion. A few items have information functions with long right tails, which means that they contribute to measuring severe motor symptoms, but we do not see corresponding long tails to the left. It may be that symptoms other than motor symptoms serve this function better, which is outside the scope of this article.

To estimate the progression of motor symptoms in Parkinson's disease, we used a longitudinal GLMM model. As this model was defined, we treated all patients as one group with individual model parameters that belonged to the same bivariate normal distribution. An alternative would have been to treat each cohort as a separate group with its own distribution of random effects. This would have been a natural choice if we had believed that the groups were clearly distinct. We believe that either approach can be theoretically justified.

As noted in \cite{arrington_performance_2020}, we need to be aware that the interpretation of the latent trait can shift slightly when we change the selection of items. This is the case when we remove items and recalculate the information function with regard to a latent trait based on the remaining items. With that methodology, the information of an item typically increases when the number of items in the selection is reduced, since each remaining item will have a larger influence on the measured construct \citep{arrington_performance_2020}. With our methodology, however, we retained the item parameters calculated from the full set of 34 items when we reduced the number of items. The interpretation of the latent trait is thus not affected by the item selection.

The results showed how the advantages of more advanced methods to minimize the uncertainty dissipate as the number of items increases. The additional complexity may not be worth it if we are to select a set of 20 items. However, with an item set of 5 or 10 items, the benefit of using a more advanced method for item selection can be substantial.

\bibliographystyle{apa-good.bst} 
\bibliography{paper4_references.bib}

@book{fedorov1972theory,
  title={Theory of optimal experiments},
  author={Fedorov, Valerii Vadimovich},
  year={1972},
  publisher={Academic Press}
}

@article{nguyen1992review,
  title={A review of some exchange algorithms for constructing discrete D-optimal designs},
  author={Nguyen, Nam-Ky and Miller, Alan J},
  journal={Computational Statistics \& Data Analysis},
  volume={14},
  number={4},
  pages={489--498},
  year={1992},
  publisher={Elsevier}
}

@book{givens2012computational,
  title={Computational statistics},
  author={Givens, Geof H and Hoeting, Jennifer A},
  volume={703},
  year={2012},
  publisher={John Wiley\&{}Sons}
}

@book{reckase_multidimensional_2009,
	address = {New York, NY},
	title = {Multidimensional {Item} {Response} {Theory}},
	isbn = {978-0-387-89975-6 978-0-387-89976-3},
	url = {http://link.springer.com/10.1007/978-0-387-89976-3},
	language = {en},
	urldate = {2023-03-19},
	publisher = {Springer},
	author = {Reckase, M.D.},
	year = {2009},
	doi = {10.1007/978-0-387-89976-3},
}

@article{goetz_movement_2008,
	title = {Movement {Disorder} {Society}-sponsored revision of the {Unified} {Parkinson}'s {Disease} {Rating} {Scale} ({MDS}-{UPDRS}): {Scale} presentation and clinimetric testing results},
	volume = {23},
	copyright = {Copyright © 2008 Movement Disorder Society},
	issn = {1531-8257},
	shorttitle = {Movement {Disorder} {Society}-sponsored revision of the {Unified} {Parkinson}'s {Disease} {Rating} {Scale} ({MDS}-{UPDRS})},
	url = {https://onlinelibrary.wiley.com/doi/abs/10.1002/mds.22340},
	doi = {10.1002/mds.22340},
	language = {en},
	number = {15},
	urldate = {2025-06-23},
	journal = {Movement Disorders},
	author = {Goetz, Christopher G. and Tilley, Barbara C. and Shaftman, Stephanie R. and Stebbins, Glenn T. and Fahn, Stanley and Martinez-Martin, Pablo and Poewe, Werner and Sampaio, Cristina and Stern, Matthew B. and Dodel, Richard and Dubois, Bruno and Holloway, Robert and Jankovic, Joseph and Kulisevsky, Jaime and Lang, Anthony E. and Lees, Andrew and Leurgans, Sue and LeWitt, Peter A. and Nyenhuis, David and Olanow, C. Warren and Rascol, Olivier and Schrag, Anette and Teresi, Jeanne A. and van Hilten, Jacobus J. and LaPelle, Nancy},
	year = {2008},
	pages = {2129--2170},
}

@book{parkinsons_progression_markers_initiative_parkinsons_2024,
	title = {{PARKINSON}'{S} {PROGRESSIVE} {MARKERS} {INITIATIVE} ({PPMI}) {Data} {User} {Guide}},
	url = {https://www.ppmi-info.org/sites/default/files/docs/PPMI\%20Data\%20User\%20Guide.pdf},
	urldate = {2025-06-26},
	publisher = {Parkinson's Progression Markers Initiative},
	author = {{Parkinson's Progression Markers Initiative}},
	month = aug,
	year = {2024},
}

@article{arrington_performance_2020,
	title = {Performance of longitudinal item response theory models in shortened or partial assessments},
	volume = {47},
	issn = {1573-8744},
	url = {https://doi.org/10.1007/s10928-020-09697-x},
	doi = {10.1007/s10928-020-09697-x},
	language = {en},
	number = {5},
	urldate = {2025-06-30},
	journal = {Journal of Pharmacokinetics and Pharmacodynamics},
	author = {Arrington, Leticia and Ueckert, Sebastian and Ahamadi, Malidi and Macha, Sreeraj and Karlsson, Mats O.},
	month = oct,
	year = {2020},
	pages = {461--471},
}

@Manual{gabry_cmdstanr_2025,
  title = {cmdstanr: R Interface to 'CmdStan'},
  author = {Jonah Gabry and Rok Češnovar and Andrew Johnson and Steve Bronder},
  year = {2025},
  note = {R package version 0.9.0, https://discourse.mc-stan.org},
  url = {https://mc-stan.org/cmdstanr/},
}

@article{vehtari_rank-normalization_2021,
	title = {Rank-Normalization, Folding, and Localization: An Improved {Rˆ} for Assessing Convergence of {MCMC} (with Discussion)},
	volume = {16},
	issn = {1936-0975, 1931-6690},
	shorttitle = {Rank-{Normalization}, {Folding}, and {Localization}},
	url = {https://doi.org/10.1214/20-BA1221},
	doi = {10.1214/20-BA1221},
	number = {2},
	urldate = {2025-07-02},
	journal = {Bayesian Analysis},
	author = {Vehtari, Aki and Gelman, Andrew and Simpson, Daniel and Carpenter, Bob and Bürkner, Paul-Christian},
	month = jun,
	year = {2021},
	pages = {667--718},
}

@article{rizopoulos_ltm_2007,
	title = {ltm: An {R} Package for Latent Variable Modeling and {Item} {Response} {Analysis}},
	volume = {17},
	copyright = {Copyright (c) 2006 Dimitris Rizopoulos},
	issn = {1548-7660},
	shorttitle = {ltm},
	url = {https://doi.org/10.18637/jss.v017.i05},
	doi = {10.18637/jss.v017.i05},
	language = {en},
	urldate = {2025-07-02},
	journal = {Journal of Statistical Software},
	author = {Rizopoulos, Dimitris},
	year = {2007},
	pages = {1--25},
}

@article{samejima_estimation_1969,
	title = {Estimation of latent ability using a response pattern of graded scores},
	volume = {34},
	number = {4, Pt. 2},
	journal = {Psychometrika Monograph Supplement},
	author = {Samejima, Fumiko},
	year = {1969},
	pages = {100--100},
}

@incollection{mcculloch_chapter_2003,
	title = {Chapter 4: {Generalized} linear mixed models ({GLMMs})},
	volume = {7},
	shorttitle = {Chapter 4},
	url = {https://projecteuclid.org/ebooks/nsf-cbms-regional-conference-series-in-probability-and-statistics/Generalized-Linear-Mixed-Models/chapter/Chapter-4-Generalized-linear-mixed-models-GLMMs/10.1214/cbms/1462106064},
	urldate = {2025-07-05},
	booktitle = {Generalized {Linear} {Mixed} {Models}},
	publisher = {Institute of Mathematical Statistics},
	author = {McCulloch, Charles E.},
	month = jan,
	year = {2003},
	doi = {10.1214/cbms/1462106064},
	pages = {28--34},
}

@article{casella_introduction_1985,
	title = {An {Introduction} to {Empirical} {Bayes} {Data} {Analysis}},
	volume = {39},
	issn = {0003-1305},
	url = {https://www.tandfonline.com/doi/abs/10.1080/00031305.1985.10479400},
	doi = {10.1080/00031305.1985.10479400},
	number = {2},
	urldate = {2025-07-06},
	journal = {The American Statistician},
	author = {Casella, George},
	month = may,
	year = {1985},
	pages = {83--87},
}

@incollection{zafar_parkinson_2025,
	address = {Treasure Island (FL)},
	title = {Parkinson {Disease}},
	copyright = {Copyright © 2025, StatPearls Publishing LLC.},
	url = {http://www.ncbi.nlm.nih.gov/books/NBK470193/},
	language = {eng},
	urldate = {2025-07-06},
	booktitle = {{StatPearls}},
	publisher = {StatPearls Publishing},
	author = {Zafar, Saman and Yaddanapudi, Sridhara S.},
	year = {2025},
	pmid = {29261972},
}

@article{almahadin_parkinsons_2020,
	title = {Parkinson’s disease: current assessment methods and wearable devices for evaluation of movement disorder motor symptoms - a patient and healthcare professional perspective},
	volume = {20},
	issn = {1471-2377},
	shorttitle = {Parkinson’s disease},
	url = {https://doi.org/10.1186/s12883-020-01996-7},
	doi = {10.1186/s12883-020-01996-7},
	number = {1},
	urldate = {2025-07-08},
	journal = {BMC Neurology},
	author = {AlMahadin, Ghayth and Lotfi, Ahmad and Zysk, Eva and Siena, Francesco Luke and Carthy, Marie Mc and Breedon, Philip},
	month = nov,
	year = {2020},
	pages = {419},
}

@article{le_when_2021,
	title = {When national drug surveys "take too long": {An} examination of who is at risk for survey fatigue},
	volume = {225},
	issn = {1879-0046},
	shorttitle = {When national drug surveys "take too long"},
	doi = {10.1016/j.drugalcdep.2021.108769},
	language = {eng},
	journal = {Drug and Alcohol Dependence},
	author = {Le, Austin and Han, Benjamin H. and Palamar, Joseph J.},
	month = aug,
	year = {2021},
	pmid = {34049103},
	pmcid = {PMC8282613},
	pages = {108769},
}

@article{ueckert_improved_2014,
	title = {Improved utilization of {ADAS}-cog assessment data through item response theory based pharmacometric modeling},
	volume = {31},
	issn = {1573-904X},
	doi = {10.1007/s11095-014-1315-5},
	language = {eng},
	number = {8},
	journal = {Pharmaceutical Research},
	author = {Ueckert, Sebastian and Plan, Elodie L. and Ito, Kaori and Karlsson, Mats O. and Corrigan, Brian and Hooker, Andrew C. and {Alzheimer’s Disease Neuroimaging Initiative}},
	month = aug,
	year = {2014},
	pmid = {24595495},
	pmcid = {PMC4153970},
	pages = {2152--2165},
}
\section*{Acknowledgements}

Data used in the preparation of this article was obtained on 2025-04-12 from the Parkinson’s 
Progression Markers Initiative (PPMI) database (www.ppmi-info.org/access-data-specimens/download-data), RRID:SCR\_006431. For up-to-date information on the study, visit www.ppmi-info.org. 

PPMI – a public-private partnership – is funded by the Michael J. Fox Foundation for Parkinson’s Research 
and funding partners, including 4D Pharma, Abbvie, AcureX, Allergan, Amathus Therapeutics, Aligning 
Science Across Parkinson's, AskBio, Avid Radiopharmaceuticals, BIAL, BioArctic, Biogen, Biohaven, 
BioLegend, BlueRock Therapeutics, Bristol-Myers Squibb, Calico Labs, Capsida Biotherapeutics, Celgene, 
Cerevel Therapeutics, Coave Therapeutics, DaCapo Brainscience, Denali, Edmond J. Safra Foundation, Eli 
Lilly, Gain Therapeutics, GE HealthCare, Genentech, GSK, Golub Capital, Handl Therapeutics, Insitro, Jazz 
Pharmaceuticals, Johnson \& Johnson Innovative Medicine, Lundbeck, Merck, Meso Scale Discovery, Mission 
Therapeutics, Neurocrine Biosciences, Neuron23, Neuropore, Pfizer, Piramal, Prevail Therapeutics, Roche, 
Sanofi, Servier, Sun Pharma Advanced Research Company, Takeda, Teva, UCB, Vanqua Bio, Verily, Voyager Therapeutics, the Weston Family Foundation and Yumanity Therapeutics.

\end{document}